\documentclass[aps,twocolumn,groupedaddress,showpacs,amssymb]{revtex4}
\input epsf

\begin{document}

\bibliographystyle{apsrev}

\title{Deformation and Flow of a Two-Dimensional Foam Under Continuous
Shear}

\author{G. Debr\'egeas, H. Tabuteau}
\author{J.-M. di Meglio}
\thanks{Also at Universit\'e Louis Pasteur and Institut Universitaire de
France}
\affiliation{Institut Charles Sadron, CNRS UPR 022, \\
 6, rue Boussingault, 67083 Strasbourg Cedex, France \\} 

\date{\today}

\begin{abstract}
We investigate the flow
properties of a two-dimensional aqueous foam submitted
to a quasistatic shear in a Couette geometry.
 A strong localization of the flow (shear banding) at the edge of
the moving wall is evidenced, characterized by an exponential decay
of the average tangential velocity. Moreover, the analysis of the
rapid velocity fluctuations reveals
self-similar dynamical structures consisting of clusters of bubbles
rolling as rigid bodies. To relate the instantaneous (elastic)
and time-averaged (plastic) components of the strain, we develop a
stochastic
model where irreversible rearrangements are 
activated by local stress fluctuations originating from the rubbing of
the wall. This
model gives a complete description of our observations and is also
consistent
with data obtained on granular shear bands by other groups.
\end{abstract}


\maketitle

Cellular materials, such as foams, concentrated emulsions, slurries, or
granular materials, exhibit rheological properties that cannot be 
understood within the scope of
standard solid or liquid mechanics \cite{lequeux,langer_liu,Nagel}. 
For such systems, thermal energies are orders of magnitude 
lower than the typical
energy required to relax the structural arrangements of their components;
 under small forces, the material remains
trapped in a metastable configuration and exhibits a solid-like behavior.
 When submitted to a large enough stress however, it 
can be driven through a
sequence of new metastable configurations, giving rise to a macroscopic
flow. But the resulting flow field may still differ a lot from what 
would be expected for a molecular liquid.

Dry sand slowly flowing down an hourglass provides a simple example
 of such abnormal flow behaviors: the flow splits into a
plug-like central region and a strongly sheared
thin layer at the wall - a few particles wide -
 where most of the dissipative process occurs \cite{pouliquen}. This
spontaneous localisation of the strain in narrow regions of the material (the
so-called shear bands)  can be observed in many other situations such as
shear,  surface, or convective   flows for instance 
\cite{behr2d, mri_chicago, komatsu, convection}. Shear banding  actually
controls most of the practical situations one has to face in soil mechanics
and industrial handling of grains, and is also relevant to pyroclastic flows
in geology (for a review on granular matter see \cite{deGennes_review}).  This
question has recently received a lot of attention from
physicists, both theoretically and experimentally   \cite{roux,jossdeb}, but a
clear picture has not emerged yet.

By contrast, the possibility of shear
banding in foams has been mostly ignored in the literature, and 
numerical or theoretical studies usually assume shear flows in foams to be
uniform \cite{langer_liu,durian1}. The assumption that shear banding
is unique to granular matter can be misleading because it suggests
that some peculiar aspects of granular flows, such as solid
friction, particle rotation or dilatancy, are required to derive a shear
band model.

In this Letter, we report the formation of shear bands in aqueous
foams.  We believe that foams may shed light on the dynamics of granular 
systems by evidencing the minimal set of ingredients
needed to get shear banding. To that extent, foams constitute a much
simpler model than granular systems since the basic bubble/bubble
interactions which
control the mechanical properties of the material are well known:
elastic
(stored) energy is related to an increase of the total interfacial area
when
the bubbles are distorted whereas dissipated energy is associated with
neighbors swapping events (\(T_{1}\) processes) inducing flows in
the liquid films and vertices
(for a review on foams, see \cite{livre_weaire}).     

In order to probe the microdynamics of the foam, one
needs to track the trajectory of each bubble during shearing. 
Since 3-D foams are inherently diffusive to light, we used a
2-D model foam - a monolayer of bubbles - submitted to a
continuous slow shear in a Couette geometry. The setup
was composed of an inner shearing wheel and an outer ring (of respective
radius
\( R_{0}=71\, \)mm and \( R_{1}=122\, \)mm) confined between two
transparent
plates separated by a \( 2\, \)mm gap. To produce the
foam, the cell was first hold vertically and partially
filled
with a controlled volume of soap/water solution. Bubbles were formed by
blowing nitrogen gas through
two small injection holes at different flow rates until the resulting
foam reached
the top of the cell. Once set horizontally, 
the foam rapidly attained a uniform wetness characterized by
its liquid fraction \( 0.01<\phi <0.3 \) (Fig. 1). This foaming
procedure was chosen because it produces
bidisperse disordered foams and therefore eliminates crystallisation.
The mean diameters within each of the two populations
of bubbles were of the order of \( 2 \) and  \( 2.7\, \)mm,
with
a mean deviation of \( 0.2\, \)mm. These bubbles
were large enough compared to the gap height so that they would not
overlap  and form a truly two-dimensional foam. To define a
bubble scale, we measured the mean distance \( d \) between first
neighbors
in the foam. In all experiments, \( d \) lay between \( 2.1 \) and \(
2.5\,  \)mm
so that the gap between the wheel and the ring could accommodate from \(
20 \)
to \( 25 \) rows of bubbles. The distance \( d \) was evaluated several
 times during
the experiment and found to be almost constant. Coarsening would
eventually
lead to a growth of the biggest bubbles at the expense of the smallest ones,
but
over a longer time. We also checked the absence of shear induced size
segregation
that might have occurred during the experiment.

\begin{figure}[tp]
\centerline{ \epsfxsize=8.5truecm
\epsfbox{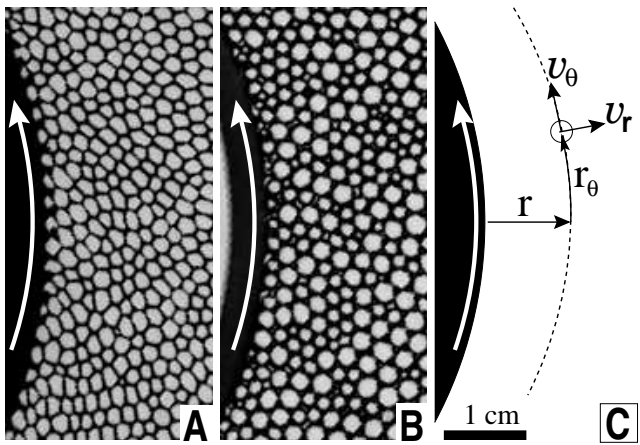}}
 \caption{Close-up frames of dry and wet 2-D foams under continuous
shear.
The shearing wheel appears in black. Notice the teeth shaped edge which
eliminates slippage of the first row of bubbles.
Arrows indicate the shearing direction. (\textbf{A}) A dry bidisperse
foam (\protect\( \phi =0.05\protect \)),
showing deformed polygonal cells. (\textbf{B}) A wet bidisperse foam
(\protect\( \phi =0.20\protect \)).
The bubbles are circular and undeformed. (\textbf{C}) System of
coordinates used to analyze the
flow field.}
 \label{figure1}
 \end{figure}

Shearing was induced by rotating the inner wheel at constant velocity \(
V_{wheel} \)
using a stepper motor. To avoid slippage at the wheel and the ring,
their sides
were teeth shaped so that the
first and last rows of bubbles would remain irreversibly
attached to the walls. To eliminate transient
effects, we ran the
experiment a full round before taking data. The motion of 1000 to 1500
bubbles was then 
recorded using a CCD digital camera positioned over the setup.
In a typical experiment, 3000 images were taken corresponding
to a total displacement of \( 600\, d \) of the wheel edge. The apparent
centers of mass of the bubbles 
were subsequently tracked by image analysis (IDL
software). To reduce the effect of the viscous friction between
the bubbles and the confining plates, we restricted our study to
quasistatic
flows. We focused on average velocity measurements as a probe of shear
rate
dependence: we found that in the range \( 0<V_{wheel}<0.7\,
\)mm.s\(^{-1}\), the
velocity profiles were similar apart from an overall scale factor. All
experiments
were performed in the quasistatic regime at \( V_{wheel}=0.25\,
\)mm.s\(^{-1}\). In the following, all velocities and distances
are normalized by \( V_{wheel} \) and \( d \) respectively. We note  \(
\omega_{0}=V_{wheel}/d \) the characteristic frequency of the shear.

\begin{figure}[tp]
\centerline{ \epsfxsize=8.5truecm
\epsfbox{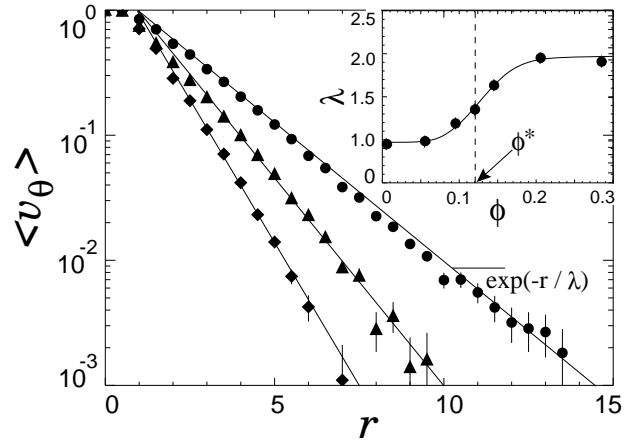} }
\caption{\protect\small 
Average tangential velocity profiles for different
liquid fraction ( $ \blacklozenge $: \protect\( \phi =0.05\protect \), $
\blacktriangle $: \protect\( \phi =0.12\protect \),
{\large $ \bullet $}: \protect\( \phi =0.20\protect \)), showing
exponential decays with
a width \protect\( \lambda (\phi )\protect \). The small plateau for
$0<r<1$ corresponds to the first row of bubbles being attached to the
shearing wheel. Inset : dependence of 
\protect\( \lambda \protect \) on liquid fractions \protect\( \phi
\protect \) (the line is just a guide for the eyes).
Two plateaus can be distinguished on either side of
\protect\( \phi ^{*}\simeq 0.12\protect \), which separate
deformed and undeformed bubbles regimes.
\label{figure2}}
\end{figure}

Figure 2(A) shows the decay of the average tangential velocity \(
\langle v_{\theta }(r) \rangle \)
with the distance \( r \) to the shearing wheel for different liquid
fractions
\( \phi  \) (see Figure 1(C) for variables definition). Averaging was
performed
over the tangential coordinate \( r_{\theta } \) and time \( t \),
yielding
smooth and reproducible profiles, although the instantaneous flow is
strongly
intermittent. The reduced velocity is found to approach \( 1 \) at \(
r\rightarrow 0 \)
confirming the absence of slip at the edge of the wheel. At larger \( r
\),
the profiles exhibit an exponential decay :
\begin{equation}
\label{exp_decay}
\langle v_{\theta }(r) \rangle \sim \exp(-r/\lambda )
\end{equation}
\noindent with a width \( \lambda  \) depending on \( \phi  \). The
curve \( \lambda  \)
versus \( \phi  \), presented on Figure 2(B), shows two plateaus at low
and high volume fraction. The transition between these two regimes
occurs around
\( \phi ^{*}=0.12 \) which qualitatively  marks
the limit between dry foams with polygonal bubbles for \( \phi <\phi
^{*} \) and wet foams with undeformed bubbles for \( \phi >\phi ^{*} \).
In both cases, the rapid decay of the mean velocity over a few bubble
diameters, establishes the existence of shear banding in foams. The exponential
shape of the velocity profile, observed in all experiments, appears as a
robust feature which was also observed in comparable experiments performed
on 2-D granular materials \cite{behr2d,pouliquen, bocquet}.

\begin{figure}[tp]
\centerline{ \epsfxsize=5truecm
\epsfbox{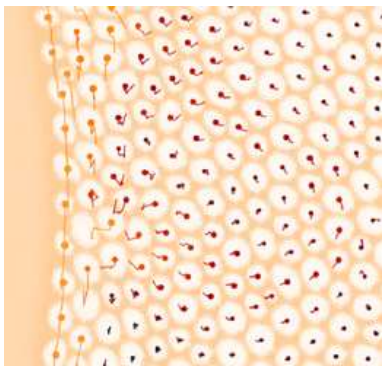} }
\caption{
Video frame of the foam with the position of the bubbles centers and their
trajectories over the last 20 seconds.
This time period corresponds to a total displacement of one bubble
diameter
for the first row (or equivalently for the inner disc edge). The dots
size and lines
color reflect the total distance travelled by the 
bubbles revealing a large rotating cluster.}
\label{figure3}
\end{figure}

Beyond these time averaged profiles, the present setup allows 
measurements
of the short timescale fluctuations of the bubbles velocities. A mere
observation
of the video sequences reveals brief oscillations of clusters of bubbles
 of various radial extension, rotating together as rigid bodies as shown
 in Figure 3. These
dynamical structures are ephemeral and disappear after the wheel edge has
moved by roughly one bubble diameter (this was checked by measuring time
 correlations of the velocity which we found decay to \( 0 \)
 in a time of the order of \(1/\omega_{0}\)). 
To quantitatively probe these coherent moves,
we studied the spatial correlations of the instantaneous velocity field.
We
focused on the radial component \( v_{r} \) which has a zero time average 
 and therefore gives a better signal to noise ratio (qualitatively,
similar results
were found when using \( v_{\theta }-\langle v_{\theta } \rangle \)
instead of \( v_{r} \)).
Figure 4(A) shows the correlation function \( g_{r}(\Delta r_{\theta })=
\langle v_{r}(r,r_{\theta}).v_{r}(r,r_{\theta }+\Delta r_{\theta })\rangle/
 \langle v_{r}^{2}(r)\rangle\)
 for different values
of \( r \) from \( 1 \) to \( 10 \), at a volume fraction \( \phi =0.20
\).
Regardless of \( r \), \( g_{r} \) decreases with \(
\Delta r_{\theta } \)
from \( 1 \) to a negative value then slowly relaxes to \( 0 \). The
length \( \xi (r) \) for which \( g_{r} \) reaches \( 0 \) defines a
typical correlation
length of the velocity field at a distance \( r \). In Figure 4(B), \(
r_{\theta } \) has been rescaled
by \( r \). All the curves then collapse on a single one,
which
demonstrates a linear increase of \( \xi (r) \) with the radial distance
\( r \): \( \xi (r)=\alpha r \).
Motivated by the observation of oscillating clusters, we modeled the
flow
field as a superimposition of rotating blocks of bubbles of various
sizes (see
figure caption for details) which allowed us to obtain a good fit of the
master curve (Figure 4(B)). Similar results
were obtained at all volume fractions, but the coefficient \( \alpha  \)
was found to decrease with \( \phi  \).

\begin{figure}[tb]
\centerline{ \epsfxsize=8.5truecm
\epsfbox{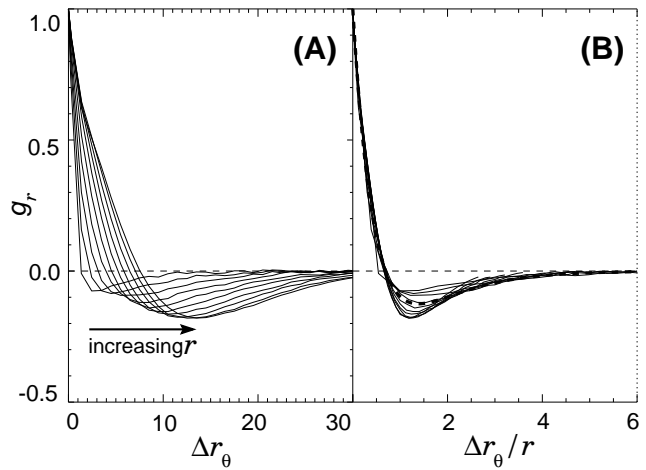} }
\caption{ 
(\textbf{A}): Spatial correlations of the radial velocity for
different radial
distances \protect\( r\protect \) from 1 to 10 (\protect\( \phi
=0.15\protect \)). 
(\textbf{B}):  All
correlations can be collapsed on a single curve when plotted versus 
the tangential
distance rescaled by \protect\( r\protect \). To fit this master
curve, we assume that the radial velocity \protect\(
v_{r}(r,r_{\theta })\protect \)
at different \protect\( r\protect \) comes from the rotation of clusters
 of mean lateral extension \protect\( \xi (r)\protect \)
proportional to \protect\( r\protect \) (\protect\( \xi (r)=\alpha
r\protect \)).
The radial velocity field within a cluster of 
size \protect\( \mu \protect \)  is idealized
by a two step function (\protect\( v_{r}=\epsilon
v_{0}\protect \)
for \protect\( -\mu <r_{\theta }<0\protect \) and \protect\(
v_{r}=-\epsilon v_{0}\protect \)
for \protect\( 0<r_{\theta }<\mu \protect \) with \protect\( \epsilon
=\pm 1\protect \)).
We postulate a statistical distribution of \protect\( \mu \protect \)
around \protect\( \xi (r)\protect \) by assuming that the
fraction
of clusters of size larger than \protect\( \mu \protect \) is
\protect\( \exp\left( -\frac{{\textstyle \mu }}{{\textstyle \xi
(r)}}\right) \protect \).
The resulting correlation function (dotted line) then reads:
\( g_{r}(\Delta r_{\theta })=\left( 1-\frac{{\textstyle \Delta r_{\theta
}}}{{\textstyle \alpha r}}\right) \exp\left( -\frac{{\textstyle \Delta
r_{\theta }}}{{\textstyle \alpha r}}\right)  \).}  
\label{figure4}
\end{figure}

It is crucial to note that the velocity field which yields the measured
correlation
functions mainly corresponds to reversible movements of the bubbles
centers,
and thus probes the elastic deformation of the foam rather than the
plastic flow.
The quantity \( \sqrt{<v_{r}^{2}>}/<v_{\theta }> \) provides a good
estimate of the ratio
of reversible to irreversible moves of the bubbles occurring upon
shearing. This
quantity is larger than \( 1 \) beyond the first attached row of
bubbles, and gets
larger than \( 10 \) beyond the fifth row. The correlation
measurements thus
reveal that the instantaneous \textit{stress} field is spatially
correlated. 

This peculiar characteristic of the foam deformation field can
actually be
understood under the scope of linear elasticity, by simply modeling
the foam as an isotropic elastic medium. 
During the initial loading, a uniform mean
stress
\( \overline{\sigma } \) develops in the material (we neglect the radial
geometry
since the wheel radius is much larger than the shear band width). In the
steady
state, this uniform stress persists in average but is locally modulated
by a
fluctuating stress field \( \Delta \sigma (t) \) of mean value \( 0 \)
associated
with the continuous rubbing of the wheel teeth. At each location \(
(r=0,\, r_{\theta }=i) \)
on the wheel edge, the foam is indeed submitted to a localized
perturbative stress  
\( \Delta \sigma _{i}(t) \) of variance \( s_{0}=\langle \Delta \sigma
_{0}^{2}\rangle \)
varying at a frequency \( \omega_{0} \) which elastically propagates into the
material.
These multiple noise sources add up to produce a stress fluctuation at a
position
\( (r,r_{\theta }) \) of amplitude (see for instance \cite{livre_johnson}):
\begin{equation}
\label{delta_sigma}
\Delta \sigma _{r}(r_{\theta },t)\simeq \sum_{i} \frac{\Delta \sigma
_{i}(t)}{\sqrt{r^{2}+(r_{\theta }-i)^{2}}}
\end{equation}
\noindent Assuming the noise sources to be uncorrelated (\( \langle
\Delta \sigma _{i}(t) \Delta \sigma _{j}(t) \rangle=0 \)
for \( i\neq j \)), the resulting stress coherence length, at a distance $r$, 
takes the form 
 \( \xi (r)=\alpha r \), in agreement with our experimental findings. 
We interpret the different observed values for \( \alpha \) as 
a signature of the anisotropy of the foam due to its initial loading.
This is consistent with the observation of large values
of \( \alpha  \)
for the driest foams where the largest uniform deformation is first produced.
 From Eq. \ref{delta_sigma}, we are also able
to compute
the variance of the fluctuating stress \( \Delta \sigma _{r}(t) \) as: 
\begin{equation}
\label{sofr}
<\Delta \sigma _{r}^{2}>=s(r)\sim \frac{s_{0}}{\alpha r}
\end{equation}

At this point, we wish to relate the fluctuating stress field
\( \Delta \sigma _{r}(t) \) to the measured average flow profile
presented
before, by taking into account the plastic property of the foam. Our
approach
mostly follows a model proposed by Pouliquen and
Gutfraind \cite{pouliquen}
based on Eyring's activated process theory \cite{eyring} to describe
chute flows
of granular materials. In the present description, the variance of the
local stress fluctuation $s(r)$ plays the role 
of a temperature allowing plastic flow to occur. The moving boundary
acts as
a ``hot wall'', exciting internal deformation modes of the material in
the
form of self-similar rotating clusters. When the  
fluctuating stress overcomes a certain yield stress \( \sigma _{y} \),
the structure plastically yields. The
yielding rate in the material by unit of time and space thus writes : \(
\omega = \omega_{0} P(\sigma >\sigma
_{y}) \)
where \( P(\sigma ) \) is the density probability of stress. The stress
at
a distance \( r \) is a sum of  \(\sim r \) random variables (see
Eq. \ref{delta_sigma})
so that \( P(\sigma (r)) \) is a gaussian distribution centered on \(
\overline{\sigma } \)
of variance \( s(r) \). Using Eq. \ref{sofr}, the yielding rate at a distance
\( r \) reads:
\begin{equation}
\label{omega}
\omega (r)=\omega_{0} P(\sigma (r)>\sigma _{y})=\omega_{0} \left( 1-{\rm
erf}\sqrt{\frac{r}{\lambda }}\right)
\end{equation}
with
\begin{equation}
\label{lambda_alpha}
 \lambda =\frac{1}{\alpha } \, \frac{2 s_{0}}{\left( \sigma
_{y}-\overline{\sigma }\right) ^{2}}
\end{equation}
Each failure increments the average velocity gradient by \( 1 \) in
reduced unit, so that
the constitutive equation for the flow writes $
 \frac{\textstyle \partial \langle v_{\theta }(r) \rangle}
{\textstyle \partial r}\sim -
\omega (r) / \omega_0$
with the boundary conditions \( \langle v_{\theta }(0) \rangle =1 \) and
\( \langle v_{\theta }(\infty )\rangle=0 \). 
A very good approximate function to the integral of \( \omega (r) \) is
given
by a pure exponential so that 
$ \langle v_{\theta }(r) \rangle \simeq \exp(-r/\lambda ) $.

This model gives the right form of the tangential velocity decay. 
An outcome of the model is contained in Eq. \ref{lambda_alpha} which
relates the
\( \alpha  \) coefficient to the plastic flow decay length \( \lambda  \) in
the foam. Experimentally, we found \( \alpha \lambda =1.02\pm 0.08 \) (\(
\lambda  \)
ranging from 1 to 2), which would suggest that \( s_{0}/(\sigma
_{y}-\overline{\sigma })^{2} \)
is independent of the liquid fraction \( \phi  \). This quantity is
indeed
self-adjusted via \( \overline{\sigma } \) in the transient regime to
allow
enough rearrangements to occur. 

In conclusion, we have observed shear banding in dry and wet
2-D foams under continuous slow shear and we have probed the associated
elastic deformations of the foam,  characterized by brief, collective
oscillations of self-similar blocks of bubbles.  We have developed a
stochastic model which relates
the plastic flow to the stress fluctuations experienced by the foam. The
main characteristics of the flow  (rapid decay of the average velocity over a
few bubbles, large velocity fluctuations)  are very similar to what is
commonly observed in granular systems,  suggesting that the proposed mechanism
could remain valid for granular systems. As already mentioned, the predicted
exponential velocity decay has been observed in various 2-D granular shear
bands \cite{pouliquen, behr2d, komatsu}. Moreover, the velocity profile in
3-D has been shown to obey a gaussian decay in the limit of
disordered and  non spherical grains \cite{mri_chicago}. This
functional form  for the velocity profile immediately follows from the
modification  of  Eq. \ref{sofr} in 3-D which then writes   \( \textstyle
s(r)\sim \frac{\textstyle s_{0}}{\textstyle \alpha^{2} r^{2}} \) yielding a
gaussian decay for \( \langle v_{\theta } \rangle \).

\begin{acknowledgments}
We are grateful to D. Mueth for helping us with the image analysis.
We wish to thank A. Kabla, S. Roux 
and C. Josserand for stimulating discussions.
\end{acknowledgments}

\end{document}